\documentclass[doublecol]{epl2} 

\title{Molecular-orbital representation of generic flat-band models}

\author{T. Mizoguchi and Y. Hatsugai}

\institute{                    
Department of Physics, University of Tsukuba - 1-1-1 Tennoudai, Tsukuba, Ibaraki 305-8571, Japan
}

\abstract{
We develop a framework to describe a wide class of flat-band models, 
with and without a translational symmetry, by using ``molecular orbitals" introduced in the prior work
(HATSUGAI Y. and MARUYAMA I., \textit{EPL}, \textbf{95}, (2011) 20003).
Using the molecular-orbital representation, we shed new light on the 
band-touching problem between flat and dispersive bands. 
We show that the band touching occurs as a result of collapse, or the linearly dependent nature, of molecular orbitals.
Conversely, we can gap out the flat bands by modulating the molecular orbitals so that they do not collapse,
which provides a simple prescription to construct models having a finite energy gap between flat bands and dispersive bands.}

\usepackage{ulem}
\usepackage{graphicx}
\usepackage{bm}
\usepackage{braket}
\usepackage{mathtools}
\usepackage{amsfonts}
\usepackage{amsmath}

\newcommand{\mmat}[4]
           {\left(\begin{array}{cc}
               #1 & #2\\
               #3 & #4
               \end{array}\right)}

\begin{document}

\maketitle
\textbf{Introduction.-}
Flat-band models are a fertile ground for exotic phenomena in condensed-matter physics. 
For fermionic systems,
it is a source of a
ferromagnetism in the presence of the interactions~\cite{Lieb1989,Mielke1991,Mielke1991_2,Tasaki1992,Tasaki1998,Tanaka2006,Katsura2010}. 
In the magnetic systems, the flat band of the magnetic mode 
in the low-energy indicates the massive degeneracy,
resulting from the frustrated nature~\cite{Reimers1991,Garanin1999,Isakov2004}.  
Recently, the combination of flat bands and topology has also attracted considerable interests~\cite{Katsura2010,Tang2011,Sun2011,Neupert2011,Sheng2011,Pal2018}. 

So far, lots of efforts have been devoted to seek the simple tight-binding Hamiltonians 
having flat bands,
such as a Lieb lattice and related models,~\cite{Lieb1989,Tasaki1992,Tasaki1998,Sutherland1986,Vidal1998,Misumi2017},
line graphs~\cite{Mielke1991,Mielke1991_2,Katsura2010}, and partial line graphs~\cite{Miyahara2005}.
It was found in those lattices that the quantum interference arising from the geometry of the lattices
gives rise to localized eigenstates which correspond to the flat bands. 
For example, in a Kagome lattice, a simple form of the localized eigenstate is given by 
assigning the staggered weight on the elementary \lq \lq loops"~\cite{Bergman2008}.

The construction of localized eigenstates is elegant, but strongly relies on the geometry of the lattice. 
This makes it puzzling to consider the models beyond the conventional models with nearest-neighbor (NN) hoppings,
e.g., in the presence of disorders and/or the farther-neighbor hoppings. 
In this letter, we present an opposite view of the flat-band models, 
namely, we discuss how to describe \textit{dispersive} bands in the flat-band models.
Extending the notion introduced in prior works~\cite{Hatsugai2011,Hatsugai2015}, 
we show that the dispersive bands 
in a wide class of flat-band models are spanned by the non-orthogonal basis functions 
consist of a small number of sites
which are referred to as \lq \lq molecular orbitals (MOs)". 
Then, the flat bands can be viewed as a complement of that space of MOs,
and they are enforced to have zero energy if the Hamiltonian is written only by MOs.
In other words, the flat-band models, with generic hoppings including farther-neighbor hoppings, can be written by MOs.  
Even in the presence of disorders, where the band picture is not available, 
the emergence of massively degenerate eigenstates is guaranteed as far as the Hamiltonian is purely composed of MOs. 

To demonstrate the usefulness of the MO-representation, 
we revisit the band-touching problem, or, counting the degeneracy of massively degenerated states.
In the literature, this problem is discussed in terms of \lq \lq topological" nature of the lattices,
namely, the number of the degenerate modes is associated with 
that of ``loops" in the lattice~\cite{Bergman2008};
recently, a novel perspective was introduced with respect to the singularity of the Bloch wave function in the momentum space~\cite{Rhim2019}.
In the scheme of the MO-representation, 
the equivalent conclusion is obtained by counting the number of the linearly independent MOs;
more precisely, the number of the linearly independent MOs subtracted from the total number of degrees of freedom gives 
{\it the minimum of the degeneracy.} 
Further, the merit of the MO-representation is that, 
one can easily find how to lift the degeneracy between flat bands and dispersive bands.
To be more specific, we modulate the MOs so that all the MOs become linearly independent. 
We demonstrate this procedure through some concrete examples, after introducing the generic formulation. 

\textbf{Formulation.-}
The construction of flat-band models 
on the basis of MOs is conceptually similar to the ``cell construction" introduced by Tasaki~\cite{Tasaki1992,Tasaki1998}.
MO-representation was applied to several models such as generalized pyrochlore models~\cite{Hatsugai2011} 
and the minimal model for silicene~\cite{Hatsugai2015}.
Here we present a general formulation of the MO-representation applicable to both 
periodic and disordered systems. 
For simplicity, we consider the case where the flat band(s) has zero energy. 
We emphasize that the MO representation is applicable to the cases with multiple flat bands with different energies, too.
In such cases, 
the Hamiltonian
subtracted by each flat-band energy
can be written by MOs; see Ref.~\cite{Hatsugai2015} as an example. 

Let us consider a lattice model with $N$ sites for spinless fermions:
\begin{eqnarray}
{\cal H}  &= \sum_{i,j=1}^M{C}_i ^\dagger h_{ij} {C}_j, \label{eq:ham}
\end{eqnarray}
where $C_i$ is an annihilation operator of a MO $i$, ($i=1,\cdots,M$) as
\begin{eqnarray}
  C_i
  = \psi_i ^\dagger \bm{c}, 
  \end{eqnarray}
  with
$\bm{c} = (c_1, \cdots, c_N)^{\mathrm{T}}$, and
$\psi_i =(\psi_{1,i}, \cdots, \psi_{N,i})^{\mathrm{T}}$.

It should be noted that 
the MOs are not necessarily orthogonal to each other,
i.e., an anti-commutation relation $ \{ C_i,  C^{\dagger}_j \}  
= \delta_{i,j}$ does not necessarily hold. 

Using MOs, the Hamiltonian of Eq. (\ref{eq:ham}) can be written 
in an original fermion basis as 
\begin{eqnarray}
  {\cal H}  &= \bm{c}  ^\dagger H \bm{c},
 \nonumber \\ 
  H &= \sum_{i,j=1}^M\psi_i h_{ij}\psi_ j ^\dagger
  = \Psi h \Psi ^\dagger,
\end{eqnarray}
where $h$ is an $M\times M$ matrix and 
\begin{eqnarray} 
  \Psi &= (\psi_1,\cdots,\psi_M),
\end{eqnarray}
is an $N\times M$ matrix.

Now, using a simple formula for $N\times M$ and $M\times N$ matrices 
$A_{NM}$ and $B_{MN}$ (see footenote 
  \footnote{
It follows from 
$  \det_N(I_N +A_{NN}B_{NN})
=   \det_N(I_N +B_{NN}A_{NN}  )$ for generic matrices $A_{NN}, B_{NN}$ 
when $A_{NN}=(A_{NM},O_{N,N-M})$ and
    $B_{NN}=\left(\begin{array}{c}B_{MN}\\ O_{N-M,N}\end{array}\right)$.
}
)
\begin{eqnarray}
  \det_N(I_N +A_{NM}B_{MN})
  &=   \det_M(I_M +B_{MN}A_{NM}  ),
\end{eqnarray}
we obtain 
\begin{eqnarray}
  \det _N(\lambda I_N-H )
  &=&   \det _N(\lambda I_N-\Psi h\Psi ^\dagger )
  \nonumber \\
  &=& \lambda ^N\det_N(I_N- \lambda  ^{-1} \Psi h\Psi ^\dagger )
  \nonumber \\
  &=& \lambda ^N\det_M(I_M- \lambda  ^{-1}  h\Psi ^\dagger \Psi)  
  \nonumber \\
  &=& \lambda ^{N-M}\det_M(\lambda I_M- h\Psi ^\dagger \Psi). 
\end{eqnarray}
Then if $N>M$, there are $N-M$ $(>0)$ zero-energy eigenstates.
When applying the above argument to the Hamiltonian matrix in the momentum space, 
one obtains flat bands.
One may also apply it to the disordered systems as well.

\textbf{Condition for additional zero modes.-}
Additional zero modes, or band touchings for translationally invariant systems, appear when
  \begin{eqnarray}
    \det _M h &= 0, \label{eq:cond1}
\end{eqnarray}
or
\begin{eqnarray}
    \Psi_{i_1,\cdots,i_M} =0,\ ( ^\forall  i_1<\cdots<i_M) \label{eq:cond2}
  \end{eqnarray}
  where
\begin{eqnarray}
  \Psi_{i_1,\cdots,i_M} &=\det_M \
  \left(\begin{array}{ccc}
    \Psi_{i_1,1} &\cdots&    \Psi_{i_1,M} \\
&\ddots& \\     
    \Psi_{i_M,1} &\cdots&    \Psi_{i_M,M} \\
  \end{array}\right)
\end{eqnarray}
is an $M$-th minor of $\Psi$.
It follows from a relation  
$\det_M h\Psi ^\dagger \Psi =\det_M h  \sum |\Psi_{i_1,\cdots,i_M}|^2$.
This is a condition for the dimension of
the projected linear space
spanned by $M$ column vectors of $\Psi h$ is less than $M$.
It implies that the projected nonzero energy bands are collapsed.
Intuitively, the condition of Eq. (\ref{eq:cond2}) corresponds to the case where the MOs are not linearly independent of each other,
thus the existence of the additional zero modes can be predicted by counting the number of linearly independent MOs.
It should be noted, however, that the counting of linearly independent MOs cannot capture the number of additional zero modes. 
More precisely, we can predict from the conditions of 
Eqs. (\ref{eq:cond1}) and (\ref{eq:cond2}) whether the reduction of the rank of $h\Psi ^\dagger \Psi$
occurs, but can not predict the rank itself;  
see the case of $\alpha$-${\cal T}_3$ model for the non-trivial example, where the number of additional zero modes increases due to a symmetry.
So, the counting of the linearly independent MOs tells us {\it the minimum of the number of additional zero modes.}
In the following, 
we elucidate, by using the specific models, how to construct the MOs
and provide a simple view of the origin of additional zero modes,
as well as a method of how to erase that touching. 

\textbf{Chiral symmetric case.-}
Before looking at specific examples, we present the generic argument on a typical case, namely, chiral symmetric systems with sublattice imbalance.
The Hamiltonian takes a form of $H=\mmat{O_{AA}}{D_{AB}}{D_{AB} ^\dagger }{O_{BB}}$,  where
$D_{AB}$ is
an $N_A\times N_B$ matrix ($N_A>N_B$).
It satisfies $\{H, \Gamma \}=0$ with $\Gamma ={\rm diag}\, (1_A,-1_B)$.
This $H$  is actually rewritten by a multiplet $\Psi$,
that is a set of $M=2N_B$ MOs 
with its total dimension $N=N_A+N_B$, as 
\begin{eqnarray*}
    \Psi &=   \mmat{D_{AB}}{O_{AB}}{O_{BB}}{1_B}, \ \
    h= \mmat{O_{BB}}{1_B}{1_B}{O_{BB}}
  \end{eqnarray*}
By the general argument,
it implies that there exist $N-M=N_A-N_B$ zero modes.
This is well known
  \footnote{
   One may check it directly as
  \begin{eqnarray*}
    \det\mmat{\lambda 1_A}{-D_{AB}}{-D_{AB} ^\dagger }{\lambda 1_B}
    =
    \det\mmat{\lambda 1_A }{D_{AB}}{O_{BA}}{\lambda 1_B -\lambda ^{-1}  D_{AB} ^\dagger D_{AB}}
    \\    = \lambda ^{N_A-N_B}\det\nolimits_B
    (\lambda ^2 1_B -  D_{AB} ^\dagger D_{AB}).
  \end{eqnarray*}
    }.
Further, additional zero modes appear when
$\det  h \Psi ^\dagger \Psi=(-)^{N_B}\det_{N_B} D_{AB} ^\dagger D_{AB}=0$.
  
\textbf{Example 1:Kagome-lattice models.-}
\begin{figure}
\includegraphics[width = 0.9\linewidth]{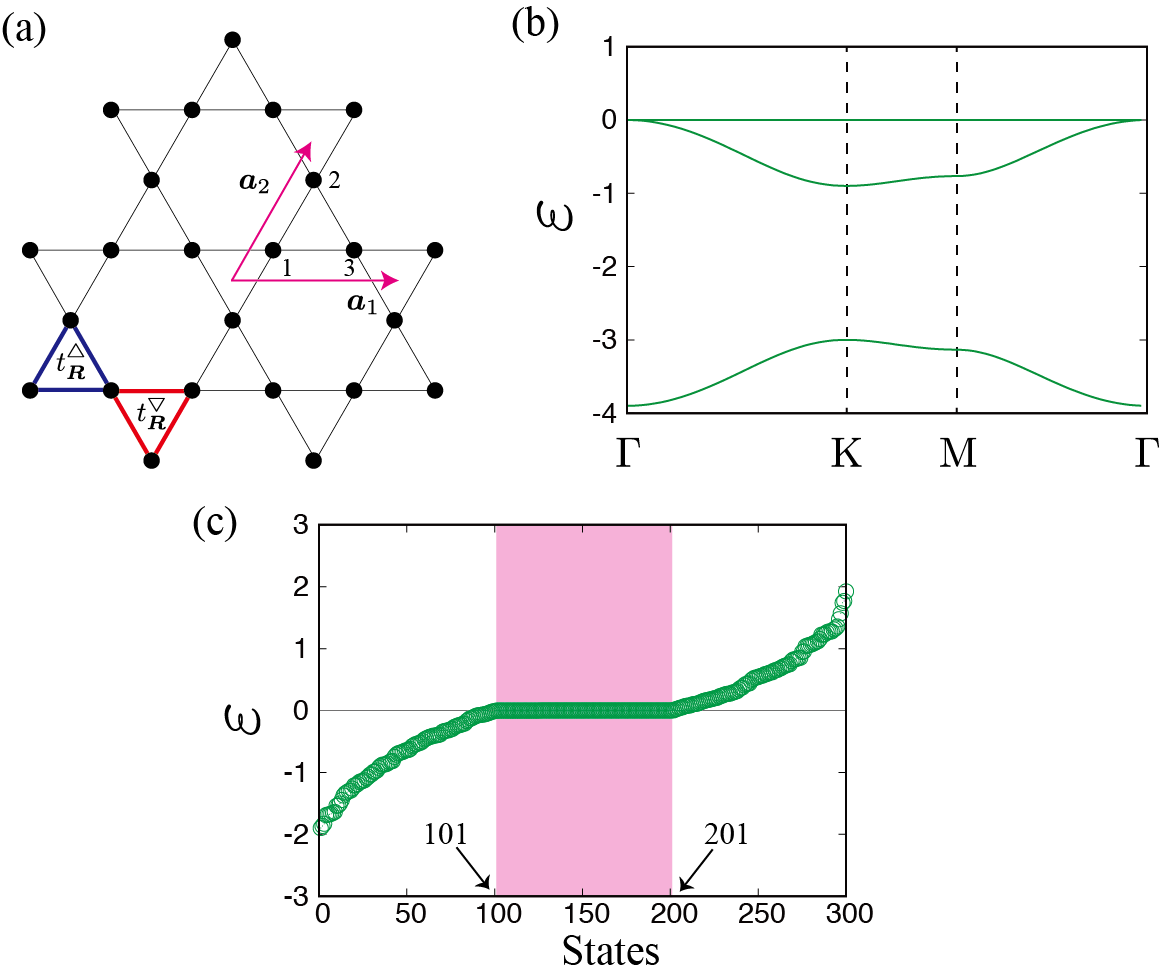}
\caption{(a) A bond-disordered Kagome model. 
The positions of the sublattices and primitive vectors are indicated in the figure.
(b) The band structure for the translationally invariant case, with $t^{\bigtriangleup} = -1.0, t^{\bigtriangledown} =- 0.3$. 
(c) The energy spectrum for the disordered model. The red shade represents the zero-energy states.}
\label{fig_kagome}
\end{figure}
We first consider a tight-binding model on a Kagome lattice with $L \times L$ unit cells [Fig.~\ref{fig_kagome}(a)].
The Hamiltonian considered here is written as 
\begin{equation}
\mathcal{H} =   \sum_{i,j\in \bigtriangleup_{\bm{R}}} t^{\bigtriangleup}_{\bm{R}} c^{\dagger}_i c_j 
+ \sum_{i,j \in  \bigtriangledown_{\bm{R}} } t^{\bigtriangledown}_{\bm{R}} c^{\dagger}_i c_j, \label{eq:hami_kagome}
\end{equation}
where $i$ and $j$ denote sites on a Kagome lattice, $\bigtriangleup_{\bm{R}}$ and $\bigtriangledown_{\bm{R}}$ denote the upward and downward triangles 
on the unit cell $\bm{R}$, respectively, and 
$t^{\bigtriangleup}_{\bm{R}}$ and $t^{\bigtriangledown}_{\bm{R}}$ are transfer integrals. 

The translationally-invariant case, i.e.,
$t^{\bigtriangleup}_{\bm{R}} = t^{\bigtriangleup}$, $t^{\bigtriangledown}_{\bm{R}} = t^{\bigtriangledown}$ for all $\bm{R}$,
is known as a breathing Kagome model~\cite{Ezawa2018,Xu2017,Kunst2018,Mizoguchi2018}. 
In this model, the band touching between flat and dispersive bands occurs~\cite{Bergman2008,Hatsugai2011,Mizoguchi2018} [Fig.~\ref{fig_kagome}(b)]. 
In the disordered case,
i.e. $t^{\bigtriangleup}_{\bm{R}}$ and $t^{\bigtriangledown}_{\bm{R}}$ are chosen randomly, 
we can not adopt the band picture. 
Nevertheless, the analogous phenomenon of the band touching, namely, 
the additional degeneracy of the zero modes, occurs. 
It should be noted that the present choice of the randomness is not likely to realize in solid-state systems, since we need a fine tuning
between the on-site potential and the NN hopping. 
Nevertheless, the present model is of fundamental importance 
because it will bring a novel perspective of disordered flat bands~\cite{Goda2006,Nishino2007,Chalker2010,Bilitewski2018}.
Indeed, the present model is quite different from the conventional disordered models with random on-site potentials,
in that the degeneracy of zero modes is exactly retained as we will show below.

In what follows, we pursue the origin of an additional zero mode by using the MO-representation. 
We define the MO on each triangle: 
\begin{equation}
C_{\bm{R}, \bigtriangleup } = c_{\bm{R} ,1 } + c_{\bm{R} ,2} + c_{\bm{R}, 3 }, \label{eq:kagome_mo_1}
\end{equation} 
\begin{equation}
C_{\bm{R}, \bigtriangledown} = c_{\bm{R} + \bm{a}_1 ,1} + c_{\bm{R} +\bm{a}_1- \bm{a}_2 ,2} + c_{\bm{R}, 3 }, \label{eq:kagome_mo_2}
\end{equation} 
then, the the Hamiltonian of Eq.~(\ref{eq:hami_kagome}) is written as 
\begin{eqnarray}
\mathcal{H} = & \sum_{\bm{R}}  t^{\bigtriangleup}_{\bm{R}}   C^{\dagger}_{\bm{R},  \bigtriangleup } C_{\bm{R},  \bigtriangleup }
 +  t^{\bigtriangledown}_{\bm{R}}  C^{\dagger}_{\bm{R},\bigtriangledown }  C_{\bm{R},  \bigtriangledown }.  \nonumber \\ \label{eq:ham_kagome_mo}
\end{eqnarray} 
Since the number of sites is $3 \times L \times L$ and the number of MOs is $2 \times L \times L$,
the model has at least $L \times L$ zero modes, from the aforementioned argument. 
In addition, the model has an additional zero mode. 
We confirm this numerically: In Fig. \ref{fig_kagome}(c), we show the energy spectrum for $L= 10$ 
with $t^{\bigtriangleup}_{\bm{R}},\hspace{1mm} t^{\bigtriangledown}_{\bm{R}} \in [-1/2,1/2]$, chosen randomly. 
In fact, we find that $101 \hspace{.5mm} (=10 \times 10 +1)$ modes out of $300$ have zero-energy. 
\begin{figure}
\includegraphics[width = 0.9\linewidth]{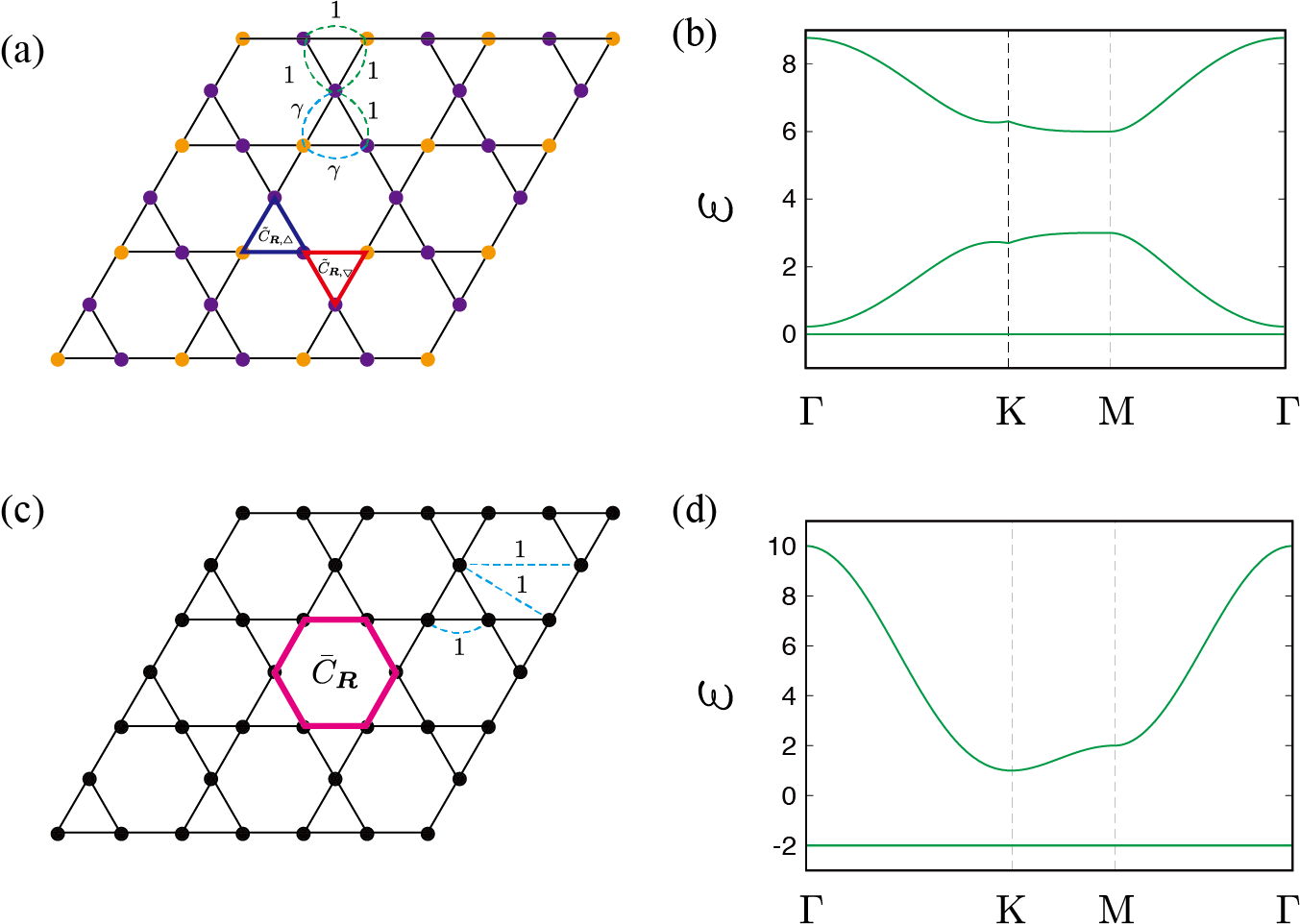}
\caption{
Kagome-lattice models with a gapped flat band.
(a) A model in Ref. \cite{Bilitewski2018}, and (b) its band structure for $\gamma = 2$. 
The orange and purple sites have an on-site potential with $\gamma^2 + 1$ and $2$, respectively.  
Blue and red triangles denote $\tilde{C}_{\bigtriangleup}$ and $\tilde{C}_{\bigtriangledown}$, respectively. 
(c) A model in Ref. \cite{Bergman2008}, and
(d) its band structure.
A hexagonal plaquette colored in magenta denotes $\bar{C}$. }
\label{fig_kagome_2}
\end{figure}
The origin of the additional zero mode can be understood as follows.
Since a set of all upward triangles covers all the sites on a Kagome lattice, and so does that of all downward triangles,
we have 
\begin{eqnarray}
\sum_{\bm{R}} C_{\bm{R}, \bigtriangleup }  = \sum_{\bm{R}}  C_{\bm{R}, \bigtriangledown}  = \sum_{\bm{R} , a = 1,2,3} c_{\bm{R}, a}. \label{eq:kagome_mo_relation}
\end{eqnarray} 
This implies that one of the MOs is written by the linear combination of the others, e.g., 
\begin{eqnarray}
C_{\bm{R}_0, \bigtriangleup }  =\sum_{\bm{R}} C_{\bm{R}, \bigtriangledown}  
- \sum_{\bm{R} \neq \bm{R}_0 } C_{\bm{R}, \bigtriangleup }. 
\end{eqnarray} 
Therefore, the dimension of the space spanned by the MOs 
is not $2 \times L \times L$ but $2 \times L \times L -1$, 
which gives rise to one additional zero mode. 

The above argument indicates that the flat band in the models written by the MOs of (\ref{eq:kagome_mo_1}) and (\ref{eq:kagome_mo_2}) can not be gapped out.  
However, in the previous works, several models on a Kagome lattices 
in which the flat band does not touch the dispersive band are proposed~\cite{Bergman2008, Bilitewski2018}. 
This means that those models are written by the different MOs, 
and in the following we show the explicit forms of those MOs, one-by-one. 

First, let us consider the model in Ref.~\cite{Bilitewski2018}, shown in Fig.~\ref{fig_kagome_2}(a). 
The MOs describing this model are obtained by modifying (\ref{eq:kagome_mo_1}) and (\ref{eq:kagome_mo_2}), slightly,
\begin{eqnarray}
\tilde{C}_{\bm{R}, \bigtriangleup } &=& \gamma c_{\bm{R} ,1 } + c_{\bm{R} ,2} + c_{\bm{R}, 3 }, \nonumber \\
\tilde{C}_{\bm{R}, \bigtriangledown} &=& c_{\bm{R} + \bm{a}_1  ,1} + c_{\bm{R} + \bm{a}_1- \bm{a}_2 ,2} + c_{\bm{R}+ \bm{a}, 3 }, 
\end{eqnarray} 
with $|\gamma| \neq 1$.
Note that $\tilde{C}_{\bm{R}, \bigtriangleup }$ and $\tilde{C}_{\bm{R}, \bigtriangledown}$ are linearly independent
of each other, due to the following reason. 
Since $\tilde{C}$'s are defined in a translationally-invariant manner, 
one can perform the Fourier transformation
\begin{eqnarray}
\tilde{C}_{\bm{k},\bigtriangleup} &=& \psi^{\dagger}_{\bm{k},\bigtriangleup,} \bm{c}_{\bm{k}}, \nonumber \\
\tilde{C}_{\bm{k},\bigtriangledown} &=& \psi^{\dagger}_{\bm{k},\bigtriangledown} \bm{c}_{\bm{k}},
\end{eqnarray} 
with $\bm{c}_{\bm{k}} = (c_{\bm{k},1} ,c_{\bm{k},2}, c_{\bm{k},3})^{\mathrm{T}}$, 
$\psi^{\dagger}_{\bm{k}, \bigtriangleup} = \left( \gamma, 1,1 \right)$, 
and $\psi^{\dagger}_{\bm{k},\bigtriangledown} = \left( e^{ i \bm{k}\cdot \bm{a}_1 } , e^{i \bm{k}\cdot (\bm{a}_1-\bm{a}_2)  }, 1 \right)$. 
Clearly, $\{\tilde{C}_{\bm{k}, \bigtriangleup/\bigtriangledown}, \tilde{C}^\dagger_{\bm{k}^\prime,\bigtriangleup/\bigtriangledown} \} = 0$ if $\bm{k} \neq \bm{k}^\prime$. 
Furthermore,  
$\tilde{C}_{\bm{k}, \bigtriangleup} $ and $\tilde{C}_{\bm{k}, \bigtriangledown}$ 
are linearly independent of each other, or,  
$\psi^{\dagger}_{\bm{k},\bigtriangleup}$ is not parallel to $\psi^{\dagger}_{\bm{k},\bigtriangledown}$,
because
$\psi^{\dagger}_{\bm{k},\bigtriangleup} \times \psi^{\dagger}_{\bm{k},\bigtriangledown} =  ( 1 -e^{i \bm{k}\cdot (\bm{a}_1-\bm{a}_2)  }, e^{i \bm{k}\cdot \bm{a}_1 }-\gamma  ,\gamma e^{i \bm{k}\cdot (\bm{a}_1-\bm{a}_2) } - e^{ i \bm{k}\cdot  \bm{a}_1 } ) \neq \bm{0}$, for all $\bm{k}$,
if $|\gamma| \neq 1$. 

Now, consider the Hamiltonian 
$\mathcal{H} =  \sum_{\bm{R}} \tilde{C}^{\dagger}_{\bm{R},  \bigtriangleup } \tilde{C}_{\bm{R},  \bigtriangleup }
 + \tilde{C}^{\dagger}_{\bm{R},\bigtriangledown }  \tilde{C}_{\bm{R},  \bigtriangledown }$.
Since all $\tilde{C}$'s are linearly independent as mentioned above, 
the band touching does not occur [Fig.~\ref{fig_kagome_2}(b)], as pointed out in Ref.~\cite{Bilitewski2018}.
The present method to gap out the flat band is simple in a sense that it does not require 
the search for suitable real-space texture of the hoppings to realize
the localized eigenstate on large loops~\cite{Bilitewski2018}. 
Also, it is easy to construct a model with disorders, 
as $\mathcal{H} =  \sum_{\bm{R}} t_{\bm{R}}^\bigtriangleup \tilde{C}^{\dagger}_{\bm{R},  \bigtriangleup } \tilde{C}_{\bm{R},  \bigtriangleup }
 + t_{\bm{R}}^\bigtriangledown \tilde{C}^{\dagger}_{\bm{R},\bigtriangledown }  \tilde{C}_{\bm{R},  \bigtriangledown }$.

Next, let us consider the second model, shown in Fig.~\ref{fig_kagome_2}(c), 
which contains not only the on-site and NN terms but also next NN and the third NN terms~\cite{Bergman2008}.
The MOs are completely different from (\ref{eq:kagome_mo_1}) and (\ref{eq:kagome_mo_2}). 
Namely, MOs are defined on the hexagonal loops, rather than the triangles, as 
\begin{equation}
\bar{C}_{\bm{R} } = c_{\bm{R},2} + c_{\bm{R},3} +  c_{\bm{R} + \bm{a}_1,1}  +  c_{\bm{R}+ \bm{a}_1,2} +  c_{\bm{R}+ \bm{a}_2,1} + c_{\bm{R}+ \bm{a}_2,3},
\end{equation} 
that are linearly independent of each other. 
This choice of MOs is inferred from the analogous spin model~\cite{Balents2002}. 
Then, consider the Hamiltonian $\mathcal{H} =  \sum_{\bm{R}} \bar{C}^{\dagger}_{\bm{R} } \bar{C}_{\bm{R}}$. 
This Hamiltonian has two flat bands and one dispersive band, 
which is consistent with the fact that 
the number of linearly independent MOs is $L \times L$
[Fig.~\ref{fig_kagome_2}(d)]. 

As shown above, there are various choices of the MOs even on the same lattices,
thus the presence/absence of the band touching is crudely dependent not on the lattice geometry but on the choice of the MOs. 

\textbf{Example 2: $\alpha$-${\cal T}_3$ model.-}
\begin{figure}
\includegraphics[width = 0.9\linewidth]{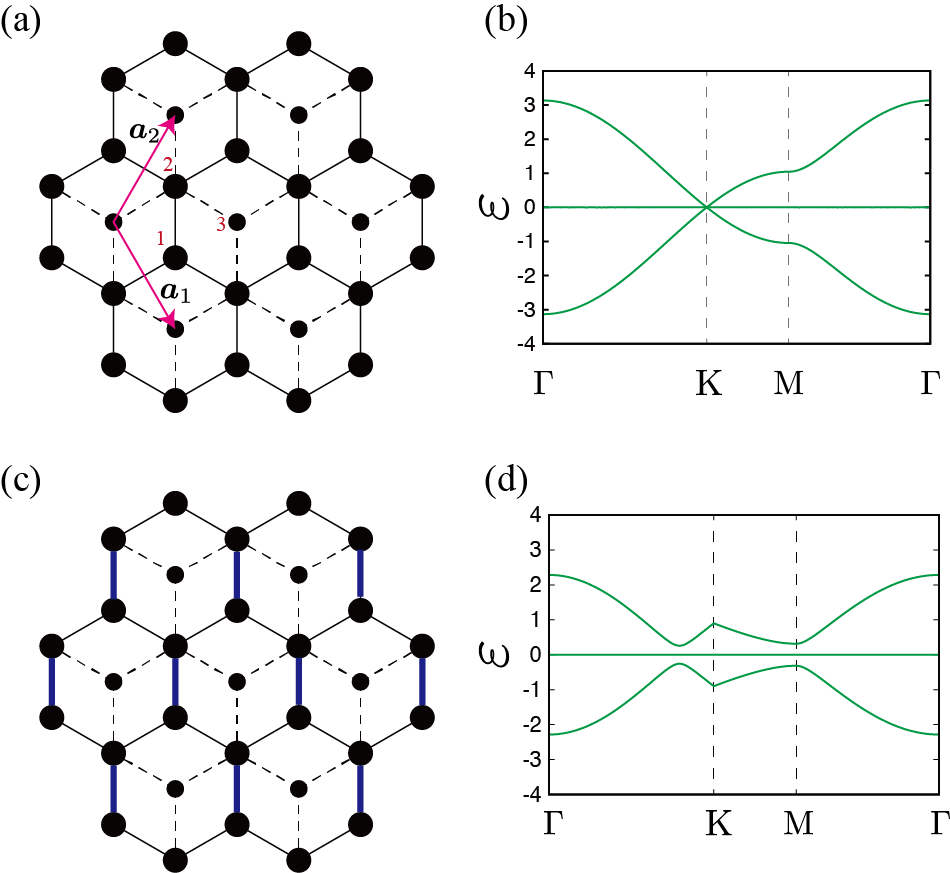}
\caption{(a) A $\alpha$-${\cal T}_3$ model.
The solid (dashed) bonds have the hopping integral 
$1$ ($\alpha$). 
The positions of the sublattices and primitive vectors are indicated in the figure.
(b) The band structures for $\alpha = 0.3$.
(c) A modulate $\alpha$-${\cal T}_3$ model. The blue bonds have the hopping $\beta$, instead of 1.
(d) The band structure for $\alpha = 0.3$, $\beta = 0.1$.
 }
\label{fig:alpha}
\end{figure}
Let us move on to the second example, namely, 
$\alpha$-${\cal T}_3$ model~\cite{Sutherland1986,Vidal1998,Raoux2014} [Fig.~\ref{fig:alpha}(b)]:
\begin{equation}
\mathcal{H} = \sum_{\bm{R}} \sum_{n =1}^3 c^{\dagger}_{\bm{R}, 1} c_ {\bm{R} + \bm{\xi}_n,2  } + \alpha c^{\dagger}_{\bm{R}, 2} c_ {\bm{R} +  \bm{\eta}_n,3} 
+ (\mathrm{h.c.}),
\label{eq:ham_alphat3}
\end{equation}
where $\bm{\xi}_1 = 0$, $\bm{\xi}_2 = \bm{a}_1$, and $\bm{\xi}_3 = -  \bm{a}_2$; 
$\bm{\eta}_1 = 0$, $\bm{\eta}_2 = - \bm{a}_1$, and $\bm{\eta}_3 =-\bm{a}_1 - \bm{a}_2$.
The model is an example of the chiral symmetric case with sublattice imbalance, 
and is known to have a flat band for arbitrary $\alpha$.
The band structure for $\alpha = 0.3$ is shown in Fig. \ref{fig:alpha}(b).
We see that the flat band, having the zero energy, touches 
the dispersive band at the K point, where the dispersive bands form a Dirac point.
Since there are two Dirac points (K and K$^\prime$ points), 
the number of zero-energy mode is $L\times L +4$, if $L$ is a multiple of three 
\footnote{
This condition is necessary in order that the K and K$^\prime$ points are included in 
the set of discretized momenta. 
}.

For this model, we define two MOs to rewrite the Hamiltonian, as 
\begin{eqnarray}
d_{\bm{R},a} &=& (c_{\bm{R},1 } + c_{\bm{R}-\bm{a}_1,1 }  + c_{\bm{R} + \bm{a}_2,1 }  ) \nonumber \\
&+& \alpha (c_{\bm{R},3} + c_{\bm{R}-\bm{a}_1,3 } + c_{\bm{R}-\bm{a}_1-\bm{a}_2 ,3 }),    \label{eq:mo_alpha1}
\end{eqnarray}
and 
\begin{equation}
d_{ \bm{R},b} = c_{\bm{R},2}.\label{eq:mo_alpha2}
\end{equation}
Notice that $d_{\bm{R},a}$ and $d_{\bm{R}^\prime,b}$ are orthogonal to each other 
since they have the weight on different sublattices.  
By using these MOs, the Hamiltonian of Eq. (\ref{eq:ham_alphat3}) can be written as
\begin{eqnarray}
\mathcal{H} = & \sum_{\bm{R}} d^{\dagger}_{ \bm{R},a} d_{\bm{R},b} +  d^{\dagger}_{ \bm{R},b} d_{ \bm{R}, a} \nonumber \\
 = & \sum_{\bm{k}} d^{\dagger}_{ \bm{k},a} d_{\bm{k},b} +  d^{\dagger}_{ \bm{k},b} d_{\bm{k},a}, \label{eq:ham_alpha_mo}
\end{eqnarray}
where $d_{\bm{k},a/b} = \sum_{\bm{R}}d_{\bm{R} ,a/b} e^{-i \bm{k} \cdot \bm{R}}$.

Now, let us examine the relation between the number of linearly independent MOs and the degeneracy of zero modes. 
At K point, where the triple band touching occurs, 
we have
\begin{eqnarray}
d_{\bm{k}_{\mathrm{K}},a} &= & \sum_{\bm{R}} e^{- i \bm{k}_{\mathrm{K}}\cdot \bm{R}} [
 (c_{\bm{R},1 } + c_{\bm{R}-\bm{a}_1,1 }  + c_{\bm{R} + \bm{a}_2,1 }  )  \nonumber \\
&+ &  \alpha (c_{\bm{R},3} + c_{\bm{R}-\bm{a}_1,3 } + c_{\bm{R}-\bm{a}_1-\bm{a}_2 ,3 })
] \nonumber \\
&= & 0.
\end{eqnarray}
Similarly, $d_{\bm{k}_{\mathrm{K}^\prime},a}$ vanishes, too. 
Thus, the number of linearly independent MOs is $2 L\times L-2$. 
However, as mentioned before, the number of zero modes is $L\times L+4$ rather than $L\times L+2$, 
meaning that there exist two additional zero modes.
The origin of these is chiral symmetry. To be specific, if we define an operator $\Gamma$ such that 
$\Gamma d_{a} \Gamma = d_{a}$, $\Gamma d_{b} \Gamma = -d_{b}$,
the Hamiltonian satisfies $\Gamma \mathcal{H} \Gamma = -\mathcal{H}$.
Consequently, $d_{\bm{k}_{\mathrm{K}},b} $ and $d_{\bm{k}_{\mathrm{K}^{\prime}},b} $ 
serve as additional zero-energy modes, although they are linearly independent of the other MOs.

How can we lift the additional degeneracy by tuning the model?
To do this, we modify $d_{\bm{R},a}$ slightly, as we did for the Kagome model:
\begin{eqnarray}
\tilde{d}_{\bm{R},a}  &=  & (\beta c_{\bm{R},1 } + c_{\bm{R}-\bm{a}_1,1 }  + c_{\bm{R} + \bm{a}_2,1 }  ) \nonumber \\
&+& \alpha (c_{\bm{R},3} + c_{\bm{R}-\bm{a}_1,3 } + c_{\bm{R}-\bm{a}_1-\bm{a}_2 ,3 }),   
\end{eqnarray}
with $|\beta| \neq 1$.
Since $\tilde{d}_{\bm{R},a}$ are linearly independent of each other, 
the Hamiltonian given as
$\mathcal{H} = \sum_{\bm{R}}
\tilde{d}^{\dagger}_{ \bm{R},a} d_{\bm{R},b} +  d^{\dagger}_{ \bm{R},b} \tilde{d}_{ \bm{R}, a}, $
does not have band-touching points [Fig.~\ref{fig:alpha}(d)]. 
The corresponding hoppings in the real space is shown in Fig.~\ref{fig:alpha}(c).
A similar method to lift the degeneracy is used in Ref.~\cite{Torma2018} 
for a Lieb lattice. 

\textbf{Summary and discussions.-}
We have shown that the MOs can express the Hamiltonian 
of a wide class of flat-band models, and that the subspace spanned by MOs corresponds to the eigenspace of dispersive modes.
Therefore, by counting the dimension of this subspace, we can count the number of flat modes as well, 
since the flat bands correspond to the co-space of MOs. 

To demonstrate its usefulness, we have studied the flat band models with and without translational symmetry,
focusing on the band-touching problem. 
From the viewpoint of MOs, 
the additional degeneracy of flat bands and dispersive bands appears
when the MOs are not linearly independent of each other. 
This indicates that, to gap out the flat band, one needs a modulation of MOs such that all the MOs become linearly independent of each other.
This offers a simple way to construct models with gapped flat bands. 

It is worth noting that, although the mathematical formulation of the MO-representation is generic 
and in principle applicable to any flat-band models\footnote{Note, however, that the MOs are not necessarily finite ranged for all flat-band models.}, 
it is not easy to search the MOs for given flat-band models systematically. 
Nevertheless, it is instructive to see that the choice of MOs for well-known flat-band models, 
namely, the NN hopping models on line graphs (e.g. a Kagome lattice and a pyrochlore lattice) 
and sublattice-imbalanced lattices (e.g. a Lieb lattice and an $\alpha$-$\mathcal{T}$-3 lattice),
is inferred from their characteristic lattice structures.  
In the former case, the natural choice is to define MOs 
on their dual lattices, where the elemental unit of those lattices,
e.g., triangles for a Kagome lattice, are placed.
In the latter case, a typical structure is that a site of ``poorer" sublattice is surrounded by sites of ``richer" sublattice.
Then, the MOs can be chosen as a poorer site itself [Eq. (\ref{eq:mo_alpha2}) for $\alpha$-${\cal T}_3$ model], 
and a linear combination of richer sites surrounding a poorer site [Eq. (\ref{eq:mo_alpha1})]. 
We emphasize that the MO representation provides a unified treatment for 
line graphs and sublattice-imbalanced lattices.

It is also worth noting that the MO representation is useful for identifying the flat-band models with farther-neighbor hoppings.
Namely, once the MOs are obtained for the NN hoppings, we can examine 
whether or not the farther-neighbor hoppings in the model can be written by the same MOs. 
For instance, the NN model on a Kagome lattice is written by the ``on-site" term of the MOs as in Eq. (\ref{eq:ham_kagome_mo}), thus 
one can easily implement the farther-neighbor hoppings retaining the flat band by introducing, e.g., ``the NN hopping term" of the MOs.
Such an implementation of the farther-neighbor hoppings 
was discussed by one of the authors in the context of tuning of the flat-band energy~\cite{Mizoguchi2018}. 
We hope that the MO-representation sheds light on the physics of flat-band models in various contexts. 

\acknowledgments
We are grateful to M. Udagawa for fruitful discussions and careful reading of the manuscript. 
This work is partly supported by Grants-in-Aid for Scientic Research, KAKENHI, JP17H06138 and JP16K13845 (YH), MEXT, Japan.


\begin{thebibliography}{0}
\bibitem{Lieb1989} 
\Name{Lieb E. H.} 
\REVIEW{Phys. Rev. Lett.}{62}{1989}{1201} 

\bibitem{Mielke1991} 
\Name{Mielke A.} 
\REVIEW{J. Phys. A: Math. Gen.}{24}{1991}{L73}
 
\bibitem{Mielke1991_2} 
\Name{Mielke A.}
\REVIEW{J. Phys. A: Math. Gen.}{24}{1991}{3311}. 

\bibitem{Tasaki1992} 
\Name{Tasaki H.} 
\REVIEW{Phys. Rev. Lett.}{69}{1992}{1608}

\bibitem{Tasaki1998} 
\Name{Tasaki H.} 
\REVIEW{Prog. Theor. Phys.}{99}{1998}{489}

\bibitem{Tanaka2006}
\Name{Tanaka A. \and Tasaki H.}
\REVIEW{Phys. Rev. Lett.}{98}{2006}{116402}

\bibitem{Katsura2010}
\Name{Katsura H., Maruyama I., Tanaka A. \and Tasaki H.}
\REVIEW{EPL}{91}{2010}{57007}

\bibitem{Reimers1991}
\Name{Reimers J. N., Berlinsky A. J. \and Shi A.-C.}
\REVIEW{Phys. Rev. B}{43}{1991}{865}

\bibitem{Garanin1999}
\Name{Garanin D. A. \and Canals B.}
\REVIEW{Phys. Rev. B}{59}{1999}{443}

\bibitem{Isakov2004}
\Name{Isakov S. V., Gregor K., Moessner R. \and Sondhi S. L.}
\REVIEW{Phys. Rev. Lett.}{93}{2004}{177204}

\bibitem{Tang2011}
\Name{Tang E., Mei J.-W. \and Wen X.-G.}
\REVIEW{Phys. Rev. Lett.}{106}{2011}{236802}

\bibitem{Sun2011}
\Name{Sun K., Gu Z., Katsura H. \and Das Sarma S.}
\REVIEW{Phys. Rev. Lett.}{106}{2011}{236803}

\bibitem{Neupert2011}
\Name{Neupert T, Santos L., Chamon C. \and Mudry C.}
\REVIEW{Phys. Rev. Lett.}{106}{2011}{236804}

\bibitem{Sheng2011}
\Name{Sheng D. N., Gu Z.-C., Sun K. \and L. Sheng} 
\REVIEW{Nat. Commun.}{2}{2011}{389}

\bibitem{Pal2018}
\Name{Pal B.}
\REVIEW{Phys. Rev. B}{98}{2018}{245116}

\bibitem{Sutherland1986}
\Name{Sutherland B.}
\REVIEW{Phys. Rev. B}{34}{1986}{5208}

\bibitem{Vidal1998}
\Name{Vidal J., Mosseri R., \and Dou\c{c}ot B.}
\REVIEW{Phys. Rev. Lett.}{81}{1998}{5888}

\bibitem{Misumi2017} 
\Name{Misumi T. \and Aoki H.}
\REVIEW{Phys. Rev. B}{96}{2017}{155137} 

\bibitem{Miyahara2005}
\Name{Miyahara S., Kubo K., Ono H., Shimomura Y., \and Furukawa N.}
\REVIEW{J. Phys. Soc. Jpn.}{74}{2005}{1918}

\bibitem{Bergman2008}
\Name{Bergman D. L., Wu C. \and Balents L.}
\REVIEW{Phys. Rev. B}{78}{2008}{125104}

\bibitem{Hatsugai2011} 
\Name{Hatsugai Y. \and Maruyama I.}
\REVIEW{EPL}{95}{2011}{20003} 

\bibitem{Hatsugai2015} 
\Name{Hatsugai Y., Shiraishi K. \and Aoki H.}
\REVIEW{New J. Phys.}{17}{2015}{025009} 


\bibitem{Rhim2019}
\Name{Rhim J.-W. \and Yang B.-J.}
\REVIEW{Phys. Rev. B}{99}{2019}{045107}

\bibitem{Ezawa2018}
\Name{Ezawa M.}
\REVIEW{Phys. Rev. Lett.}{120}{2018}{026801} 

\bibitem{Xu2017} 
\Name{Xu Y., Xue R. \and Wan S.}
arXiv:1711.09202
 
\bibitem{Kunst2018} 
\Name{Kunst F. K., van Miert G. \and Bergholtz E. J.} 
\REVIEW{Phys. Rev. B}{97}{2018}{241405(R)} 

\bibitem{Mizoguchi2018}
\Name{Mizoguchi T. \and Udagawa M.}
\REVIEW{Phys. Rev. B}{99}{2019}{235118} 

\bibitem{Bilitewski2018}
\Name{Bilitewski T. \and Moessner R.}
\REVIEW{Phys. Rev. B}{98}{2018}{235109}

\bibitem{Goda2006}
\Name{Goda M., Nishino S. \and Matsuda H.}
\REVIEW{Phys. Rev. Lett.}{96}{2006}{126401}

\bibitem{Nishino2007}
\Name{Nishino S., Matsuda H. \and Goda M.}
\REVIEW{J. Phys. Soc. Jpn.}{76}{2007}{024709}

\bibitem{Chalker2010}
\Name{Chalker J. T., Pickles T. S. \and Shukla P.}
\REVIEW{Phys. Rev. B}{82}{2010}{104209}

\bibitem{Balents2002}
\Name{Balents L., Fisher M. P. A. \and Girvin S. M.}
\REVIEW{Phys. Rev. B}{65}{2002}{224412}

\bibitem{Raoux2014}
\Name{Raoux A., Morigi M., Fuchs J.-N., Pi\'{e}chon F., \and Montambaux G.}
\REVIEW{Phys. Rev. Lett.}{112}{2014}{026402}

\bibitem{Torma2018}
\Name{T\"{o}rm\"{a} P., Liang L. \and Poetta S.}
\REVIEW{Phys. Rev. B}{98}{2018}{220511(R)}

\end{thebibliography}
\end{document}